\documentclass[prl,twocolumn]{revtex4}  		
\usepackage{graphicx}

\begin{document}

\title
{
Nature of carriers in one-dimensional correlated electron systems
}

\author{
Hiroki Tsuchiura$^{1,@}$, Masao Ogata$^{2}$, Yukio Tanaka$^{3}$
and Satoshi Kashiwaya$^{4}$
}
\affiliation{
CREST, Japan Science and Technology Corporation (JST), Nagoya 464-8603$^{1}$\\
Department of Physics, University of Tokyo, Bunkyo-ku, Tokyo
113-0033, Japan$^{2}$ \\
Department of Applied Physics, Nagoya University, Nagoya
464-8603, Japan$^{3}$ \\
NRI of AIST, Umezono, Tsukuba, Ibaraki 305-8568, Japan$^{4}$
} 

\date{\today}

\begin{abstract}
Electron densities at the edge site in one-dimensional Hubbard model
and $t$-$J$ model are studied by using the Bethe ansatz solutions
and exact diagonalization method.
It is found that the boundary is electron-attractive, or equivalently
hole-repulsive, near half-filling.
We propose a new criterion to determine whether the carriers in
a strongly correlated system are electron-like or hole-like depending
on the electronic behavior at the boundary.
\end{abstract}

\pacs{Valid PACS appear here}%

\maketitle
A suitably substituted impurity atom in condensed matter systems can
play a role as an atomic scale probe for local response of the system.
In particular in strongly correlated electron systems, the response
to an impurity offers a key to understand the nontrivial
properties of the system.
Therefore, a large number of studies have been carried out
on the impurity effects in the strongly correlated electron systems,
especially in high-$T_{c}$ superconductors.
One of the most important findings concerning the impurity effects
in high-$T_{c}$ superconductors is the appearance of local
magnetic moments around a nonmagnetic impurity, such as Zn or Li atom,
substituted for Cu in the CuO$_{2}$ plane in underdoped compounds.
\cite{alloul,mahajan,ishida2,bobroff,julien}.

Recently, we have pointed out that whether the impurity in the CuO$_{2}$
plane tends to
expel electrons or holes is crucial for the appearance of induced
moments or the local enhancement of antiferromagnetic (AF) correlation
\cite{znaf}.
If the system is in the overdoped region, it has been shown that
electrons are expelled from the neighborhood of the impurity and
thus a hole-rich region is locally formed around the impurity.
As a result, AF correlation is collapsed because the effective
exchange coupling between electrons is locally suppressed.
Conversely, if the nature of the carriers is hole-like in
the underdoped region or close to half-filling, we expect that
the holes are scattered by the impurity and then an electron-rich
region is formed around the impurity.
If this is the case, it is natural to consider that local magnetic
moments appear near the impurity since the local region approaches
half-filling.
Therefore it is very important to clarify the nature of carriers
in correlated electron systems, on which the local magnetism around
an impurity strongly depends.

In the two-dimensional systems, however, it is difficult to determine
the nature of carriers doped in the Mott insulator.
Slave-boson mean-field theory suggests that the charge degrees of freedom
can be represented as a small number of holons, however this naive picture
is uncertain because there exists the large Gauge fluctuation.
In this paper, we investigate the nature of carriers in one-dimensional (1D)
systems where the essence of strong correlation can be clarified in
exact solutions.

It is worthwhile noting here that even in 1D systems, the nature of
carriers has not been understood because there is no Hall effect.
Thermopower was suggested to clarify the charge of the carriers, but
its calculation is only phenomenological
\cite{schulz0}.
Instead of these quantities we propose that the local electron density
near the impurity gives a good quantity for determining whether the carriers
are hole-like or electron-like.
When the impurity is repulsive, it scatters the mobile carrier.
Near half-filling, we expect that the holes are scattered and the hole
density should decrease around the impurity.
Using this criterion we determine the phase diagram in the 1D Hubbard model.

When we assume the impurity scattering is in the unitarity limit,
the impurity is equivalent to the boundary in 1D systems.
We calculate the electron density at the edge site of the 1D
Hubbard model and $t$-$J$ model with boundary by using the Bethe ansatz
solutions\cite{schulz,asakawaH,frahm1,frahm2,essler,asakawatJ}
and the exact diagonalization method.

The Hamiltonian of the open Hubbard chain with boundary fields, $p_{1}$
and $p_{L}$, is given as
\begin{eqnarray}
 {\cal H}_{\rm Hub} &=& -\sum_{j\sigma}^{L-1}
 ( c^{\dag}_{i\sigma} c_{j\sigma} + {\rm h.c.} )
 + U\sum_{j=1}^{L}n_{j\uparrow}n_{j\downarrow}
\nonumber \\
 && + \mu\sum_{j=1}^{L}(n_{j\uparrow} + n_{j\downarrow})
\nonumber \\
 && - p_{1}( n_{1\uparrow} + n_{1\downarrow} )
 - p_{L}( n_{L\uparrow} + n_{L\downarrow} ) ,
\label{hamilHub}
\end{eqnarray}
following the notation in ref.~\cite{asakawaH}.
The hopping integral, $t$, has been taken as a unit of energy ($t=1$).
We describe the length of the chain by $L$, which is assumed to be
an even integer.
The average number of electrons at the edge site $\langle n_{1}\rangle$
taken with respect to the ground state $|\Phi_{0}\rangle$ 
is obtained from
\begin{equation}
 n_{1} = -\frac{\partial}{\partial p_{1}}
 \langle \Phi_{0}|{\cal H}_{\rm Hub}|\Phi_{0}\rangle .
\end{equation}

The Bethe ansatz equations for the above Hamiltonian are 
\begin{eqnarray}
 e^{ik_{j}2(L+1)} &\cdot& 
 \frac{1-p_{1}e^{-ik}}{1-p_{1}e^{+ik}}\cdot
 \frac{1-p_{L}e^{-ik}}{1-p_{L}e^{+ik}}
\nonumber \\
 &=& \prod_{\beta=1}^{M}\frac{\sin k_{j} - \lambda_{\beta} + iu}
     {\sin k_{j} - \lambda_{\beta} - iu}
\cdot\frac{\sin k_{j} + \lambda_{\beta} + iu}
     {\sin k_{j} + \lambda_{\beta} - iu} ,
\nonumber
\end{eqnarray}
\[
 \prod_{j=1}^{N}\frac{\lambda_{\alpha} - \sin k_{j} + iu}
     {\lambda_{\alpha} - \sin k_{j} - iu}
 \cdot\frac{\lambda_{\alpha} + \sin k_{j} + iu}
     {\lambda_{\alpha} + \sin k_{j} - iu}    
\nonumber 
\]
\vspace{-0.5cm}
\[
 = \prod_{\beta\neq\alpha}^{M}
\frac{\lambda_{\alpha} - \lambda_{\beta} + 2iu}
     {\lambda_{\alpha} - \lambda_{\beta} - 2iu}
\cdot\frac{\lambda_{\alpha} + \lambda_{\beta} + 2iu}
     {\lambda_{\alpha} + \lambda_{\beta} - 2iu} ,
\]
\vspace{-0.5cm}
\begin{equation}
j = 1,\cdots, N,  ~~~ \alpha=1,\cdots, M,
\end{equation}
where $u=U/4$, $N$ is the total number of electrons and
$M$ is the number of down spins \cite{asakawaH}.
In this letter, we only discuss the repulsive Hubbard model.
%
For this case it is useful to denote the negative $k_{j}$'s and
$\lambda_{\beta}$'s as
 $k_{-j} = -k_{j}$, 
 $\lambda_{-\beta} = -\lambda_{\beta}$ .
Using standard procedures, the Bethe ansatz equations for the ground
state can be rewritten as the following linear integral equations
for $\sigma_{L}^{c}(k)$ and $\sigma_{L}^{s}(\lambda)$, which represent
the densities of solutions for the real quasimomenta $k_{j}$ 
and spin rapidities $\lambda_{\beta}$:
\[
 \sigma_{L}^{c}(k|k^{+},\lambda^{+})=\frac{1}{\pi} + \frac{1}{\pi L}
  \frac{dp_{0}}{dk} ~~~~~~~~~~~~~~~~~~~~~~~~~
\]
\vspace{-0.5cm}
\begin{equation}
 + \frac{\cos k}{2\pi}\int_{-\lambda^{+}}^{\lambda^{+}}d\lambda'
  K_{1}(\sin k-\lambda')\sigma_{L}^{s}(\lambda'|k^{+},\lambda^{+}) , 
\label{sigc} 
\end{equation}
\[
 \sigma_{L}^{s}(\lambda|k^{+},\lambda^{+}) 
 = \frac{1}{\pi L}\frac{dq_{0}}{dk} ~~~~~~~~~~~~~~~~~~~~~~~~~~~~~~~~
\]
\vspace{-0.5cm}
\[
 + \frac{1}{2\pi}\int_{-k^{+}}^{k^{+}}dk'
 K_{1}(\lambda -\sin k')\sigma_{L}^{c}(k'|k^{+},\lambda^{+})
\]
\vspace{-0.5cm}
\begin{equation}
 -\frac{1}{2\pi}\int_{-\lambda^{+}}^{\lambda^{+}}d\lambda'
  K_{2}(\lambda-\lambda')\sigma_{L}^{s}(\lambda'|k^{+},\lambda^{+}) ,
\label{sigs}
\end{equation}
where
\[
 K_{1}(x) = \frac{2u}{u^{2} + x^{2}}, ~~~
 K_{2}(x) = \frac{4u}{4u^{2} + x^{2}},
\]
are the standard kernel for the Hubbard model, and
\[
 p_{0}(k)= k + \frac{1}{2}\theta_{0}(k,p_{1})
            + \frac{1}{2}\theta_{0}(k,p_{L})
            - \tan^{-1}\frac{\sin k}{u} ,
\]
\[
 q_{0}(\lambda) = 
            - \tan^{-1}\frac{\sin\lambda}{2u}, ~~~
 \theta_{0}(k,p) = 2\tan^{-1}\left(\frac{p\sin k}{1-p\cos k}\right).
\]
The values of $k^{+}$ and $\lambda^{+}$ are determined from the
conditions
\begin{eqnarray}
 \int_{-k^{+}}^{k^{+}}dk\sigma_{L}^{c}(k|k^{+},\lambda^{+})
   &=& \frac{2N+1}{L},
\nonumber \\
 \int_{-\lambda^{+}}^{\lambda^{+}}d\lambda
   \sigma_{L}^{s}(\lambda|k^{+},\lambda^{+}) &=& \frac{2M+1}{L}.
\end{eqnarray}

In one-dimensional systems with open boundaries, the first finite-size
correction for the ground state energy $E_{L}$ is of order $L^{0}$,
which is the contributions from the {\it boundary}.
Thus the ground state energy of the present model can be written as
\begin{equation}
 E_{L} = L e_{\infty}(k^{+},\lambda^{+})
         + e_{1}(k^{+},\lambda^{+}) + O(1/L)
\label{energy}
\end{equation}
with
\begin{eqnarray}
 e_{\infty}(k^{+},\lambda^{+}) &=&
   \frac{1}{2\pi}\int_{-k^{+}}^{k^{+}}dk\varepsilon_{c}(k),
    \nonumber \\
 e_{1}(k^{+},\lambda^{+}) &=&
   \frac{1}{2\pi}\int_{-k^{+}}^{k^{+}}dk\varepsilon_{c}(k)
    \frac{dp_{0}}{dk} \nonumber \\
 &+& \frac{1}{2\pi}\int_{-\lambda^{+}}^{\lambda^{+}}d\lambda
  \varepsilon_{s}(\lambda)\frac{dq_{0}}{d\lambda}
  + 1 - \frac{\mu}{2} .
\end{eqnarray}
Here, $\varepsilon_{c}(k)$ and $\varepsilon_{s}(\lambda)$ are
the dressed energies defined as
\begin{eqnarray}
 \varepsilon_{c}(k) &=& -2\cos k + \mu 
\nonumber \\
&& + \frac{1}{2\pi}\int_{-\lambda^{+}}^{\lambda^{+}}d\lambda'
  K_{1}(\sin k-\lambda')\varepsilon_{s}(\lambda) , 
\label{epsc}
\\
 \varepsilon_{s}(\lambda) &=& \frac{1}{2\pi}
  \int_{-k^{+}}^{k^{+}}dk'\cos k'K_{1}(\lambda-\sin k')\varepsilon_{c}(k')
  \nonumber \\
  && -\frac{1}{2\pi}\int_{-\lambda^{+}}^{\lambda^{+}}d\lambda'
  K_{2}(\lambda-\lambda')\varepsilon_{s}(\lambda') .
\label{epss}
\end{eqnarray}
If the energy $E_{L}$ in the thermodynamic limit is minimized for
$k^{+}=k_{0}$ and $\lambda^{+}=\lambda_{0}$, the parameters
$k_{0}$ and $\lambda_{0}$ are obtained by
\begin{equation}
 \left.\frac{\partial e_{\infty}}{\partial k^{+}}
  \right|_{k^{+}=k_{0}} = 0, ~~~
 \left.\frac{\partial e_{\infty}}{\partial \lambda^{+}}
   \right|_{\lambda^{+}=\lambda_{0}} = 0 .
\label{1stderiv}
\end{equation}
These conditions are equivalent to
\begin{equation}
 \varepsilon_{c}(k_{0}) = 0, ~~~\varepsilon_{s}(\lambda_{0}) = 0,
\end{equation}
which are the same as in the periodic-boundary case.

Now we expand $e_{\infty}$ and $e_{1}$
up to the second order in $\Delta k = k^{+} - k_{0}$ and 
$\Delta\lambda = \lambda^{+} - \lambda_{0}$.
As far as the conditions eq. (\ref{1stderiv}) are satisfied, we find
$\Delta k$ and $\Delta\lambda$ are of the order of $1/L$, and
$k^{+}$ and $\lambda^{+}$ in eq.~(\ref{energy}) can be replaced by
$k_{0}$ and $\lambda_{0}$, whose values are determined from the conditions
\begin{equation}
 \int_{-k_{0}}^{k_{0}}dk\sigma_{\infty}^{c}(k)
   = \frac{2N}{L},    ~~
 \int_{-\lambda_{0}}^{\lambda_{0}}d\lambda
  \sigma_{\infty}^{s}(\lambda) = \frac{2M}{L} ,
\label{renormsig}
\end{equation}
Here $\sigma_{\infty}^{c}(k)$ and
$\sigma_{\infty}^{s}(\lambda)$ 
are obtained by taking the limit of $L\to\infty$ 
in eqs.~(\ref{sigc}) and (\ref{sigs}).
Since $\lambda_{0}=\infty$ for the case of $N=2M$, 
$\sigma_{\infty}^{s}(\lambda)$ can be solved 
from the Fourier transformation.
Substituting $\sigma_{\infty}^{s}(\lambda)$ into the integral equation
for $\sigma_{\infty}^{c}(k)$, we obtain
\begin{equation}
 \sigma_{\infty}^{c}(k) = \frac{1}{\pi}
 + \cos k\int_{-k_{0}}^{k_{0}}dk'R(\sin k -\sin k')\sigma_{\infty}^{c}(k')
\label{siginteg}
\end{equation}
with
\begin{equation} 
 R_{u}(k) = \frac{1}{2\pi}
 \int_{-k_{0}}^{k_{0}}dx\frac{e^{-ikx}}{1+e^{2u|x|}} .
\label{Rdef}
\end{equation}

Finally to obtain $e_{1}(k_{0},\lambda_{0})$, we need $\varepsilon_{c}(k)$
and $\varepsilon_{s}(\lambda)$.
In the same way as $\sigma_{\infty}^{c}(k)$,
we have
\begin{eqnarray}
 \varepsilon_{c}(k) &=& -\cos k + \mu 
\nonumber \\
&+& \int_{-k_{0}}^{k_{0}}dk'
 R(\sin k-\sin k')\cos k'\varepsilon_{c}(k') .
\label{epsinteg}
\end{eqnarray}
As a result, the electron density at the edge site is given by
\begin{equation}
 n_{1} = 
 -\frac{\partial e_{1}}{\partial p_{1}}\biggr|_{p_{1}=0}
 = -\frac{1}{2\pi}\int_{-k_{0}}^{k_{0}}dk\varepsilon_{c}(k)\cos k.
\end{equation}
In the actual calculation, we solve eqs.~(\ref{renormsig}) and
(\ref{siginteg}) to obtain $k_{0}$ and $\sigma_{\infty}^{c}(k)$ for
each value of $U/t$ and $N/L$.
Then using the obtained $k_{0}$, we solve $\varepsilon_{c}(k)$ in
eq.~(\ref{epsinteg}) to calculate $n_{1}$.

Figure 1 shows the obtained $n_{1}$ as a function of the electron density,
$n=N/L$, of the whole system.
In the noninteracting case ($U/t=0$), $n_{1}$ is always
smaller than $n$ in less-than-half-filling case ($n<1$), which means that
the carriers are electrons.
For $n>1$, $n_{1}$ can be easily obtained from the electron-hole symmetry
and is larger than $n$, indicating that the carriers are hole-like.
As $U/t$ increases, the crossover point from $n_{1}>n$
to $n_{1}<n$ approaches $n=0.5$ from $n=1$.
If we take the limit of $U/t\to\infty$, 
the $n_{1}$-$n$ relation is just half-size of that for $U/t=0$, i.e.,
$n=2$ for $U/t=0$ corresponds to $n=1$ for $U/t\to\infty$.
This can be easily understood because the charge degrees of freedom
are expressed as a Slater determinant of spinless fermions \cite{ogata}.
Thus, $n=0.5$ means the half-filled band of the spinless fermions.

Figure 2 shows the crossover points as a function of $t/U$.
Apparently, in the region of $n\sim 1$ and large $U/t$, $n_{1}$ is larger
than $n$. 
From these results, we interpret that the holes are scattered by the
impurity (in this case, by the boundary) and thus the nature of carriers
in this region is hole-like.
This is one of the typical effects of strong correlation in the doped
Mott insulator.

\begin{figure}[htb]
\begin{center}
\includegraphics[width=6cm]{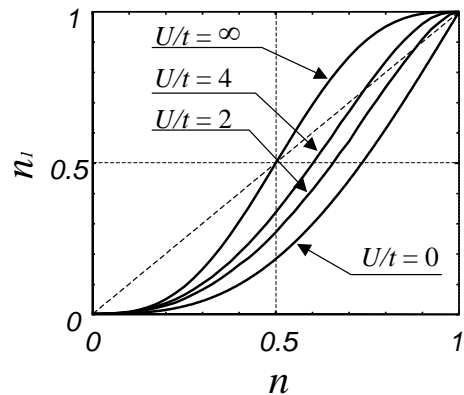}
\end{center}
\vskip -4mm
\caption{
The average number of electrons at the edge site of the open Hubbard chain
as a function of the electron density of the whole system.
A dashed line indicates $n_{1}=n$ line. 
}
\label{fig1}
\end{figure}
%
\begin{figure}[htb]
\begin{center}
\includegraphics[width=6cm]{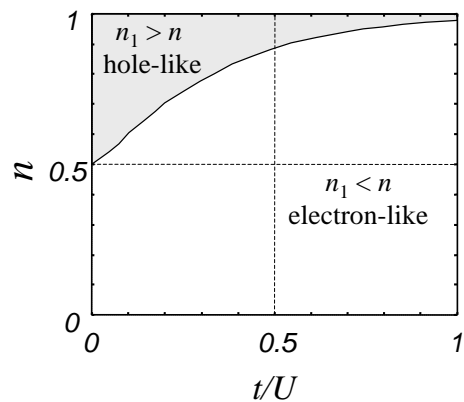}
\end{center}
\vskip -4mm
\caption{
The phase diagram in 1D Hubbard model.
The nature of carriers is hole-like in the shaded area.
}
\label{fig2}
\end{figure}

%
%

Next we study the same problem in the $t$-$J$ model.
The Hamiltonian of the open $t$-$J$ chain with boundary fields is given as
\begin{eqnarray}
 {\cal H}_{t{\mbox{-}}J} &=& -t\sum_{j=1}^{L-1}\sum_{\sigma}
  P_{G}\left( c_{j\sigma}^{\dagger}c_{j+1,\sigma} + {\rm h.c.}\right)P_{G}
\nonumber \\
 &+& J\sum_{j=1}^{L-1}\left(\mbox{\boldmath $S_{i}\cdot S_{j}$}
 - \frac{1}{4}n_{j}n_{j+1}\right)
 - \mu\sum_{j=1}^{L}n_{j} 
\nonumber \\
   &-& \frac{h}{2}\sum_{j=1}^{L}(n_{j\uparrow}-n_{j\downarrow})
   + \mu_{1}n_{1} + \mu_{L}n_{L}
\label{tjhamil}
\end{eqnarray}
following the notation in ref.~\cite{asakawatJ}.
The Bethe ansatz equations for the open $t$-$J$ model are available
only when $J/t=2$ under periodic \cite{bares1,bares2}
and open boundary conditions \cite{essler,asakawatJ}.
Following the same procedure as for the Hubbard model,
we obtain the electron density at the edge site.
For the other values of $J/t$, we calculate $n_{1}$ in the exact
diagonalization of the system with 16 sites.

Figure 3 shows the results of the $t$-$J$ model in the same manner
as in Fig. 1. 
Note that $J/t=0$ corresponds to $U/t\to\infty$ in the Hubbard model.
The exact diagonalization (ED) results show that, as $J/t$ increases,
$n_{1}$ decreases for any values of $n$ and hole-like region shrinks.
This behavior is consistent with that of the Hubbard model
in the large $U$ region.
For $J/t=2$, the ED results slightly overestimate  the value of $n_{1}$
compared to the Bethe ansatz (BA) results.
This is the finite size effect, i.e., there is no enough space where
electrons escape from the boundary in small systems.

It is interesting to compare the results for the $t$-$J$ model
with that for the Hubbard model.
For $J/t=2$, the BA result shows that $n_{1}<n$ for any values of $n$,
which means that the carrier is always electron-like.
Actually, the result for $J/t=2$ is very close to that for the free
fermion case ($U/t=0$ in Fig. 1).
The reason is as follows.
When $J/t=2$, the electron motions driven by the hopping term and by
the exchange term have the same amplitude $t=J/2$.
Therefore, electrons can move almost independently with each other
and thus a free electron-like behavior can be seen for $J/t=2$ 
\cite{yoko-ogata}.
Furthermore the result for $J/t=1$ is close to that for $U/t=4$,
which is consistent with the relation between $J$ and $U$,
i.e., $J = 4t^{2}/U$.
Even though the so-called three-site terms, which appear in the perturbation
theory with respect to $t/U$, are neglected in eq. (\ref{tjhamil}), we can
see that the nature of carriers has a strong correspondence between the
Hubbard model and the $t$-$J$ model when $J/t\alt 1$.
%
\begin{figure}[htb]
\begin{center}
\includegraphics[width=6cm]{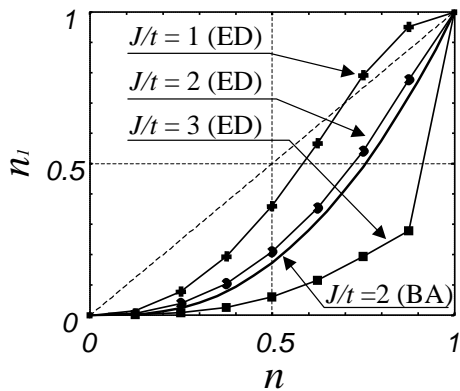}
\end{center}
\vskip -4mm
\caption{
The average number of electrons at the edge site of the $t$-$J$ model
as a function of the electron density of the whole system.
The results are obtained by using the exact diagonalization (ED)
or the Bethe ansatz equations (BA).
A dashed line indicates $n_{1}=n$ line.
}
\label{fig3}
\end{figure}


In summary, we have investigated the electron number at the edge site
of the one-dimensional Hubbard model and $t$-$J$ model by using
the Bethe ansatz solutions and the exact diagonalization method.
We propose a new criterion to determine whether the carriers in
strong correlated systems are electron-like or hole-like, which is
usually difficult in 1D because of the lack of the Hall effect.
We expect the similar behavior in 2D Hubbard model or $t$-$J$ model.
Although reliable estimation of the crossover point will be difficult,
it is natural to expect that the carriers are hole-like (or holons)
near half-filling, which causes the local magnetic moments around
impurities.
As the system approaches half-filling, the number of the hole-like
carriers decreases leading to the metal-insulator transition.

%
The authors wish to thank J. Inoue and H. Ito for their useful discussions.
Numerical computation in this work was partially carried out at
the Yukawa Institute Computer Facility, the Supercomputer Center,
Institute for Solid State Physics, University of Tokyo.
This work was partly supported by a Grant-in-Aid from the Ministry 
of Education, Science, Sports and Culture of Japan. 

%


\end{document}